\documentclass[preprint,aps]{revtex4}
\usepackage{dcolumn}
\usepackage{graphicx}
\usepackage{latexsym}
\begin{document}

\title{Can degenerate bound states occur in one dimensional quantum mechanics?}

\author{Sayan Kar}
\email{sayan@cts.iitkgp.ernet.in}
\affiliation{Department of Physics and Centre for Theoretical Studies
\\
Indian Institute of Technology, Kharagpur 721 302, WB, India}
\author{Rajesh R. Parwani}
\email{parwani@nus.edu.sg}
\affiliation{Department of Physics and University Scholars Programe, \\
National University of Singapore, Kent Ridge, Singapore.\\}

\begin{abstract}
We point out that bound states, degenerate in energy but differing in parity, may form in one dimensional quantum systems even if the potential is non-singular in any finite domain. Such potentials are necessarily unbounded from below at infinity and occur in several different contexts, such as in the study of localised states in brane-world scenarios. We describe how to construct large classes of such potentials and give explicit analytic expressions for the degenerate bound states. Some of these bound states occur above the potential maximum while some are below. Various unusual features of the bound states are described and after highlighting those that are ansatz independent, we suggest that it might be possible to observe such parity-paired degenerate bound states in specific mesoscopic systems.  
\end{abstract}

\date{March 26, 2007}
\maketitle

The purpose of ``no-go" theorems in physics is to delimit what is possible, and what is not, within a particular mathematical model of physical phenomena. Of course Nature is not obliged to respect our prejudices, so the assumptions used in proving such theorems should be well-motivated. Even then, some condition often gets slipped in, or in time forgotten, leading to confusion when counter-examples are found. In this Letter we discuss one such theorem and its possible physical consequences.

An often quoted result in quantum mechanics states that in one space 
dimension, potentials that are continuous and bounded from below cannot 
produce degenerate bound states. Elementary versions of the proof go as 
follows \cite{landau}. Consider the stationary Schrodinger equation on the open line $-\infty < x < \infty $,
\begin{equation}
\psi^{''} + k^2(x)  \psi =0 \, , \label{sch}
\end{equation}
with  $k^2(x) \equiv E-V(x)$, 
and where we assume for convenience a symmetric potential, $V(x)=V(-x)$. As the equation is real we can, without loss of generality, discuss its real solutions. For two bound states $\psi_1, \psi_2$ of the same energy, one deduces from (\ref{sch}) that the Wronskian
\begin{equation}
W \equiv \psi_2 \psi_{1}^{'} - \psi_{1} \psi_{2}^{'}  \, \label{wrons}
\end{equation}
is a constant, $W=C$. 
The constant $C$ can be evaluated at any point and for bound states the point at infinity is natural since the wavefunctions vanish there. If $C=0$ then  (\ref{wrons}) implies 
\begin{equation}
{\psi_{1}^{'} \over \psi_1 } = {\psi_{2}^{'} \over \psi_2 } \label{ratio}
\end{equation} 
and so we have $\psi_1 \propto \psi_2$: the states are not independent, there is no degeneracy. 

One way of avoiding the conclusion of the theorem was argued long ago \cite{loudon}. The expression (\ref{ratio}) is ill-defined at places where the 
wavefunction vanishes, and it was argued that if the potential was singular 
at a node of the wavefunctions then one could produce degenerate bound 
states \cite{cohen}. 
These have been studied by a number of authors, and it has also been 
suggested that some of the singular potentials approximate certain physical situations in molecular chemistry \cite{pathak} and elsewhere \cite{Wan}. However, 
it should be noted that singular potentials by themselves do not guarantee degeneracy \cite{pathak,Wan}.

Let us suppose now that the potential is non-singular in any finite domain. 
Is the proof then complete? Not quite--there is still a loophole in the above 
argument. In order for the Wronskian in (\ref{wrons}) to vanish at infinity 
for bound states we need to assume that the states have finite slopes at 
infinity. While this may seem a reasonable assumption, it is not necessarily 
true as we shall see below. The extra ingredient that is needed in the proof 
is that the  potential be bounded from below \cite{messiah}: then, if, 
asymptotically $k^2 > 0$, one has the usual continuum of scattering states
, while the bound states exist if, asymptotically, $k^2 < 0$, i.e. the energy 
$E$ lies inside a potential well. A more involved reasoning \cite{messiah} 
then shows that the bound states have the usual behaviour at infinity, with 
finite derivatives (for potentials that are finite but oscillatory at infinity one can 
have bound states in the continuum \cite{bic}, but they have finite slopes).

A physical way of understanding the relationship between the diverging slope of a bound state and unbounded potentials is as follows: If the slope diverges, the state has diverging momentum and hence diverging mean kinetic energy. Therefore the mean potential energy of the state must be negative and divergent so as to get a total mean energy, $E$, that is finite. That is, the potential must be unbounded below. 

Thus, it seems that if the potential is non-singular, then the only way to 
by-pass the theorem, and so produce degenerate bound states, is to study 
potentials that are unbounded from below at infinity. (Of course if one 
considers {\it compact} spaces, for example physics on a circle with periodic 
boundary conditions, then all states are localised and can be degenerate, 
see for example \cite{Wan}).

Bottomless potentials are neither uncommon nor necessarily pathological. For 
example the Coulomb potential is  unbounded at the origin. Yet the hydrogen 
atom has a stable ground state, prevented from collapsing by quantum 
fluctuations, as is usually argued heuristically using the uncertainty 
principle. Potentials unbounded from below have also been studied in the 
context of non-Hermetian Hamiltonians \cite{bender} which still produce real 
spectra because of a $PT$ symmetry. It has been shown that such systems may be 
mapped to a different Hermetian system \cite{mos} although the mapping is known explicitly only in a few cases \cite{bus}. 

Unbounded potentials have also appeared in the context
of localisation of fields on the brane in models with warped
extra dimensions {\cite{braneworld}}. In a recent study \cite{koleykar} 
it was found that the 
following  bottomless potential

\begin{equation}
V(x) = -(A_{1} \cosh^{2 \nu} x + A_{2}
\hspace{.1 cm} \mbox{sech}^2 x)  \, , \label{kkpot}
\end{equation}  
where $\nu >0$ and $A_{2} =  \frac{\nu}{2}(
\frac{\nu}{2} +1)$, led to opposite parity bound states with the same energy,
\begin{eqnarray}
\psi_{1}(x) =  \frac{\cos \left[\sqrt{A_{1}}  
\hspace{.1 cm} \int (\cosh x)^{\nu} dx \right]}{(\cosh x)^{\frac{\nu}{2}}} \\
\psi_{2}(x) =  \frac{\sin \left[\sqrt{A_{1}}  
\hspace{.1 cm} \int (\cosh x)^{\nu} dx \right]}{(\cosh x)^{\frac{\nu}{2}}} 
\end{eqnarray}
(see also \cite{choho} where degenerate bound states were studied for a special case of (\ref{kkpot})).
Figure 1 shows the potential, the degenerate even and odd parity states
and the energy. By tuning the parameter $A_1$ (with fixed $\nu$) 
one can have bound states inside or outside the well. 
Notice that in this example, while the wavefunctions vanish at 
infinity, their slope does not, and thus $C \neq 0$ in (\ref{wrons}). 
Potentials of such shape are sometimes called ``volcano" potentials and 
various aspects of such systems, such as the perturbation series and 
resonances, have been studied \cite{volcano}.

\begin{figure}
\centerline{
\includegraphics[width=8cm,height=5cm]{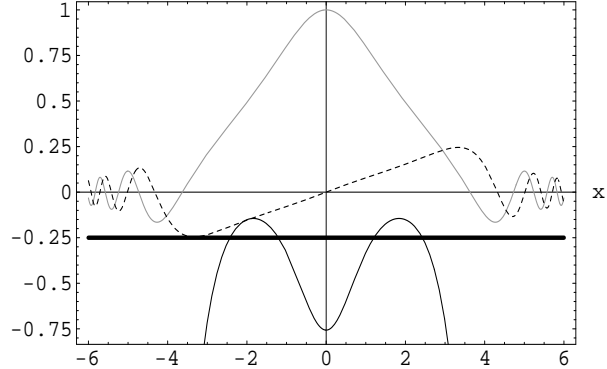}}
\caption{The potential $V(x)$ (black), the even parity wave function
(gray), the odd parity wave function (dashed) and the energy (thick black)  
for the values of the parameters $\nu = 1$ and $A_1 = \frac{1}{144}$. The 
vertical axis represents both the unnormalised amplitude of the wavefunction 
and the energy scale in appropriate units.} 
\end{figure}


The example of equation (\ref{kkpot}) is not an isolated curiosity. 
In this paper, we deduce various generic properties of degenerate bound states 
supported by non-singular, but unbounded, potentials and show how to 
explicitly construct large classes. We then discuss the possibility of 
realising them experimentally.
Note that although the unbounded systems we study generally have 
non-Hermetian boundary conditions, they do have real bound state solutions. 

Now, linear second-order differential equations of the form (\ref{sch}) are very well studied \cite{Tit}, and we quote here two results for our use. Firstly, if the potential is unbounded below then $k^2 \to + \infty$ as $x \to \pm \infty$ and so all solutions of (\ref{sch}) will be oscillatory \cite{GR}. That is, all solutions will have an infinite number of zeros. This is intuitively reasonable as in this case the equation looks asymptotically like that for a classical harmonic oscillator but with increasing frequency. 

Secondly, another well-known theorem \cite{GR} states that the zeros of two 
independent solutions of (\ref{sch}) interlace: between any two zeros of one 
solution there will be exactly one zero of the other solution. 
Again the $\sin$ and $\cos$ solutions of the classical harmonic oscillator 
equation illustrate this.

Also, as we noted above, for degenerate bound states we need (\ref{wrons}) 
to be nonzero, so at least one of the bound states, say $\psi_1$, 
must have a diverging slope at infinity. Combining this with the fact that 
the state must be oscillatory means that the zeros must get closer together 
as $x$ increases. But since the other independent state, $\psi_2$, has 
interlacing zeros with $\psi_1$, its zeros too must get closer together 
and it too must have a slope that diverges at infinity. 

We now proceed to construct classes 
of such degenerate bound states and the potentials that support them. The discussion above 
suggests we adopt the following ansatz for our pair of degenerate states

\begin{eqnarray}
\phi_{+}(x) = f(x) \cos g(x) \label{plus} \\
\phi_{-}(x) = B  f(x) \sin g(x) \label{minus}
\end{eqnarray}
where $B$ is a constant and the functions $f,g$ need to be twice 
differentiable for use in (\ref{sch}). Integrability of the wavefunctions 
is achieved by requiring
\begin{equation}
\int dx \ f^2 < \infty \, . \label{integ}
\end{equation}
Substituting (\ref{plus}),(\ref{minus}) into (\ref{wrons}) we find a 
constraint on $g$,
\begin{equation}
g(x) = \int^{x} {\gamma \over f^2(x')} \ dx'\, , \label{gg}
\end{equation}
where we have set $\gamma \equiv C/B$. For this integral to be well-defined we require $f$ to be nonzero in the domain $(-\infty,\infty)$. So we choose, 
$f(x) >0$.

The corresponding potential can be re-constructed  using (\ref{plus},\ref{gg}) in (\ref{sch}),
\begin{equation}
 V(x)-E = { f^{''} \over f} - {\gamma^2 \over f^4} \, . \label{pot}
\end{equation} 
Choosing a reference point such as $V(0)=0$, fixes the energy of the degenerate bound states. 
We can make the potential symmetric by requiring $f(x)=f(-x)$, and then we see that $\phi_{+}$ is parity even while $\phi_{-}$ is odd, that is, the degenerate states have opposite parities. 

In summary, starting with any twice-differentiable, square-integrable and 
positive function $f(x)$ defined on the real line, we have an associated 
potential given by (\ref{pot}). The potential is clearly nonsingular and 
diverges to minus infinity as $x \to \pm \infty$ due to the $\gamma/f^4$ term. Note that the rate at which the potential falls off at infinity is directly correlated with how localised the bound states (\ref{plus},\ref{minus}) are, both being determined by the function $f$. Since $V(x)=V(-x)$ the potential has an extremum at the origin. 

Choosing $f(x)={1 \over (\cosh x)^{\nu/2} }$ gives the example 
of Ref.\cite{koleykar}. There is clearly a very wide choice in constructing 
other explicit examples, such as $f(x) = \exp(-x^2)$ which also gives rise 
to exponentially localised bound states. Before we look at another example 
from this class let us discuss some features of the solutions in 
the class (\ref{plus}),(\ref{minus}) that are independent of the specific 
form of $f$. 

Since the energy is finite, so $E \equiv <H>=<T+V>$ must be finite. But 
from (\ref{plus},\ref{integ}) and (\ref{pot}) we see that $<V>$ generally 
diverges, getting large negative contributions at infinity. This means that 
the average kinetic energy, $<T>$, must likewise be divergent, getting 
large positive contributions at infinity. This is consistent with the fact 
that the state has diverging slope at infinity, meaning diverging momentum. 
Related to the diverging kinetic energy is the fact that the states (\ref{plus}),(\ref{minus}) clearly have an infinite number of nodes, see (\ref{gg}) and (\ref{integ}), whose density increases with $x$.

Let us now look in detail at another example from the above class of solutions,
\begin{equation}
f(x) = {a \over \sqrt{1 + x^2} } \label{case2}
\end{equation}
with $a$ a constant. This gives $g(x) = {\gamma \over a^2} ( { x^3 \over 3} + x )$ for use in the corresponding bound states (\ref{plus},\ref{minus}). The potential is 
\begin{equation}
V(x)-E = \frac{2 x^2-1}{\left(x^2+ 1\right )^2}-
\frac{\gamma^2}{a^4} \left(x^2+1 \right)^2  \, . \label{case2pot}
\end{equation}
In addition to the extremum at $x=0$ the potential has at most two maxima located at $x=x_{\pm}$, obtained as solutions of the algebraic
equation
\begin{equation}
\left (x_{\pm}^2+ 1 \right )^4 - \frac{a^4}{\gamma^2} \left (2 - {x_{\pm}}^2\right )=0 \, . \label{algeb}
\end{equation}
It is easy to see from a graphical analysis that real positive solutions, $x_{\pm}^2$, to (\ref{algeb}) exist,  and hence the maxima exist, only if $\gamma^2 < 2 a^4$ : otherwise the bound states occur above the convex potential! When the well forms, the height of the
potential at the barriers is given by
\begin{equation}
V(x_{\pm}) -E =  \frac{3 ({x_{\pm}}^2- 1)}{ (x_{\pm}^{2} +1)^2 } \, . \label{barr}
\end{equation}
To see whether the bound states are inside or outside the well, set $z=x_{\pm}^2 >0$ and rewrite (\ref{algeb}) as a condition on $\gamma$,
\begin{equation}
\gamma^2 =  {a^4(2 - z) \over  (z^2 +1)^4}  \, \, , \label{cond}
\end{equation}
the right-hand-side of which is a decreasing function of $z$. Thus we see again that we need $\gamma^2 < 2 a^4$ for a well to form. Furthermore (\ref{barr}) 
shows that $z=1$ is a critical value: for $z<1$ the bound states are outside 
the well and they move inside when $z>1$. Using this in (\ref{cond}) 
shows that even after the well forms, the bound states remain outside 
if $\gamma^2 > a^4/16$.  

The $\gamma^2=0$ limit is interesting as it corresponds to $C=0$ in 
(\ref{wrons}) and hence no degeneracy. What happens is that as $\gamma \to 0$, 
the oscillations of both states occur further out, and since the parity odd 
state has the $\sin (\gamma ...)$ factor, it goes to zero in any finite 
domain. Eventually at strictly $\gamma=0$, the parity odd state drops out of 
the spectrum and the parity even state becomes non-oscillatory, $\phi_{+} 
\to f(x)$, representing the ground state of the potential (\ref{case2pot}) with $\gamma=0$. Since the $\gamma=0$ potential goes to a finite constant at infinity, which is conventionally choosen to be zero, so its ground state, $\phi= f$, is a zero energy bound state.

\begin{figure}
\centerline{
\includegraphics[width=7cm,height=5cm]{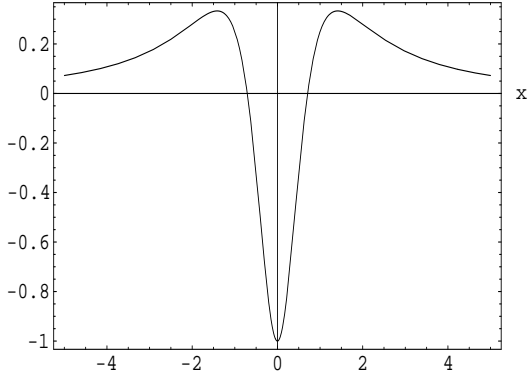}}
\caption{The potential ($V(x)-E$) in Eqn. (13) for $\gamma^2=0$.}
\end{figure}

The $\gamma^2=0$ version of (\ref{case2pot}) is a volcano potential with a 
finite bottom, Figure 2. Such potentials also occur naturally in braneworld 
scenarios \cite{grem}, the normalisable zero energy ground state corresponding 
to a stable graviton localised near the brane. Generalisations of (\ref{case2}) which lead to potentials with similar properties are discussed elsewhere \cite{kp2}.    

So far we have illustrated one class of potentials. Noting that $\sin z$ is 
related to the spherical Bessel function $J_{1/2}(z)$, one may replace the 
trigonometric functions in (\ref{plus}),(\ref{minus}) by Bessel functions of 
fractional order $J_{\pm \mu}(z)$ to obtain a different class. 
The relation (\ref{gg}) remains essentially the same but the resulting 
expression for the potential is more complicated. We discuss the Bessel function ansatz and others in \cite{kp2}. 
 
Given the fact that degenerate bound states have been shown to exist in very large classes of non-singular potentials, depending on a function $f$ with rather mild constraints,  we think that it should be possible to observe such states under laboratory conditions. In particular, advances in technology have enabled one to engineer specific mesoscopic structures \cite{meso} and we feel that these might be ideal for the purpose at hand. The generic features of the states that one is looking for are: One dimensional degenerate bound states of opposite parity supported by non-singular potentials that are unbounded below at infinity. 

The oscillatory bound states we have discussed bear some resemblance to 
von Neumann-Wigner bound states \cite{bic} that occur in the continuum. 
While the von Neumann-Wigner states are also oscillatory, they are typically 
supported in three dimensions by spatially oscillatory potentials 
( von Neumann-Wigner bound states have also been studied in low dimensions 
for specific geometries with particular boundaries, see for example, \cite{lowBIC}).  The main differences are that in our case the {\it potential is non-oscillatory} and there is a {\it pair of degenerate states.} Furthermore, in the idealised case we have discussed so far, the degenerate states can occur not only above the potential well but also inside it. 

Of course bottomless potentials, like the singular potentials discussed in \cite{loudon,cohen,pathak} are mathematical idealisations and one expects realistic potentials that approximate these to still display the main characteristics. If our bottomless potential is cut off at some large distance then we expect the bound states to be still essentially degenerate, oscillatory and appearing close to the top of the potential. Given that von Neumann-Wigner states have been detected experimentally \cite{cap}, we think that the degenerate states we have discussed in this paper may also be observable.  

We note that volcano potentials with a finite bottom have been studied before \cite{semi}, but we are unaware of any investigations which focus on possible degenerate bound states at the top of such potentials that have a large height compared to the depth of the well. 

There are many interesting questions regarding potentials of the form (\ref{pot}). For example, do they support bound states other than the ones used to construct them? Also, as noted above, in the $C \to 0$ limit the expression (\ref{pot}) becomes a bounded potential expressed in terms of a function $f$ that may be interpreted as the ground state of the system (since it is nodeless)--this suggests some connections with supersymmetric quantum mechanics \cite{susy}. We hope to return to these and other questions at a later stage \cite{kp2}.



\begin{thebibliography}{66}
\bibitem{landau}  L. D. Landau and E. M. Lifshitz, Quantum Mechanics (Pergamon Press, Oxford, 1977), Pg. 60 

\bibitem{loudon} R. Loudon, Am. J. Phys. {\bf 27}, 649 (1959). 

\bibitem{cohen} J. M. Cohen and B. Kuharetz, J. Math. Phys. {\bf 34}, 12 (1993).

\bibitem{pathak}  K. Bhattacharyya and R. K. Pathak, Int. J. Quantum
  Chem. {\bf 59}, 219 (1996); J-M. Levy Leblond and F. Balibar, 
Quantics: rudiments of quantum physics (North Holland, 1990), Pg. 357.

\bibitem{Wan} K.K. Wan, From Micro to Macro Quantum Systems, Chap. 6, (Imperial College Press, 2006).

\bibitem{messiah} A. Messiah, Quantum Mechanics, Pgs 98-106, (Dover Publications, 1999). 

\bibitem{bic} J. von Neumann and E. Wigner, Phys. Z {\bf 30}, 465 (1929);
F. H. Stillinger and D. R. Herrick, Phys. Rev. {\bf A 11},
446 (1975).

\bibitem{bender} C. Bender, Introduction to PT symmetric quantum theory, quant-ph/0501052. 

\bibitem{mos} A. Mostafazadeh, J. Math. Phys. {\bf 43}, 205 (2002);
J. Phys. {\bf A 36}, 7081 (2003).

\bibitem{bus} V. Buslaev and V. Grecchi, J. Phys. {\bf A 26}, 5541 (1993); H.F. Jones and J. Mateo, Phys. Rev {\bf D 73}, 085002 (2006).

\bibitem{braneworld} L. Randall and R. Sundrum, Phys. Rev. Lett. {\bf 83},  4690 (1999);  
C. Csaki, J. Erlich, T. J. Hollowood, and Y. Shirman, Nucl. Phys. {\bf
B 581}, 309 (2000);  R. Koley and S. Kar, Class. Quant. Grav. {\bf 22},
753 (2005); N. Barbosa-Cendejas
  and A. Herrera-Aguilar, Phys. Rev {\bf D 73}, 084022 (2006). 

\bibitem{koleykar} R. Koley and S. Kar, Phys.Lett. {\bf A 363}, 369 (2007)

\bibitem{choho} H-T. Cho and C-L. Ho, quant-ph/0606144

\bibitem{volcano} E Caliceti, V. Grecchi and M. Maioli, 
Commun. Math. Phys. {\bf 157},347 (1993); J. Zamastil, J. Cizek and L. Skala, Phys. Rev. Letts. {\bf 84}, 5683 (2000);
J. Zamastil, V. Spirko, J. Cizek, L. Skala and O. Bludsky, Phys. Rev. {\bf{A 64}},
042101 (2001); 
F. J. Gomez and L. Sesma, Phys. Letts. {\bf A 301}, 184 (2002).

\bibitem{Tit}  E.C. Titchmarsh, Eigenfunction expansions associated with 
second-order differential equations, Part I  
(2nd Edition, Oxford University Press, 1962), Pgs 92-94 and Pgs 123-125. 


\bibitem{GR} I.S. Gradshteyn and I.M. Ryzhik, Tables of Integrals, Series and Products, (Academic Press, 1980). 

\bibitem{grem} M. Gremm, Phys. Lett. {\bf B 478}, 434 (2000).

\bibitem{kp2} S. Kar and R. Parwani, in progress.

\bibitem{meso} F. Capasso, J. Faist, and C. Sirtori, J. Math. Phys. {\bf 37} 
4775 (1996); D.F. Holcomb, Am. J. Phys. {\bf 67}, 278 (1999).

\bibitem{lowBIC} P.S. Deo and A.M. Jayannavar, Phys. Rev. {\bf B 50}, 11629 (1994);
S. Longhi, quanth-ph/0612152, and references therein.



\bibitem{cap} F. Capasso, C. Sirtori, J. Faist, D. L. Sivco, S. N. G. Chu 
and A. Y. Cho, Nature {\bf 358}, 565 (1992).

\bibitem{semi} L. L. Chang, L. Esaki, R. Tsu,  Appl. Phys. Lett. {\bf 24}, 
593 (1974); 
Z. I. Alferov, Rev. Mod. Phys. {\bf 73}, 767
(2001);  T.C. Au Yeung,
Yabin Yu, W.Z. Shangguan and W.K. Chow, Phys. Rev. {\bf{B 68}}, 075316
(2003).; M. M. Nieto, Phys. Lett. {\bf B486},414 (2000).

\bibitem{susy} F. Cooper, A. Khare and  U. Sukhatme, Phys.Rept. {\bf 251} 
267 (1995).
\end{thebibliography}
\end{document}